\documentclass[conference]{IEEEtran}
\IEEEoverridecommandlockouts
\usepackage{cite}
\usepackage{amsmath,amssymb,amsfonts}
\usepackage{algorithmic}
\usepackage{graphicx}
\usepackage{textcomp}
\usepackage{xcolor}
\usepackage{multirow}
\def\BibTeX{{\rm B\kern-.05em{\sc i\kern-.025em b}\kern-.08em
    T\kern-.1667em\lower.7ex\hbox{E}\kern-.125emX}}
\begin{document}

\title{Quantum Organisational Readiness Levels\\
\thanks{Funding for this research was provided by Nationaal Regieorgaan Prak\-tijkgericht Onderzoek SIA (https://regieorgaan-sia.nl/): The Taskforce for Applied Research, part of the Netherlands Organisation for Scientific Research. The funder had no role in research design, data collection and analysis, decision to publish, or preparation of the manuscript.}
}

\author{\IEEEauthorblockN{1\textsuperscript{st} Marten Teitsma}
\IEEEauthorblockA{
\textit{Amsterdam University of Applied Sciences}\\
Amsterdam, Netherlands \\
m.teitsma@hva.nl}
\and
\IEEEauthorblockN{2\textsuperscript{nd} Iftikhar Ahmed}
\IEEEauthorblockA{
\textit{Capgemini}\\
Berlin, Germany \\
iftikhar.ahmed@capgemini.com}
\and
\IEEEauthorblockN{3\textsuperscript{rd} Julian van Velzen}
\IEEEauthorblockA{
\textit{Capgemini}\\
Utrecht, Netherlands \\
julian.van.velzen@capgemini.com}
}

\maketitle

\begin{abstract}
Setting out a path to use quantum computing within a company is not as straightforward as the implementation of classical ICT-projects. The technology is fundamentally different and not mature yet, which makes the development and use uncertain, non-linear and more complex. Being also a potential disruptive technology makes it for a company important to be aware of the possible business value generated by using quantum computing and be prepared for taking this technology in production. In this article we present Quantum Organisational Readiness Levels to determine the degree of readiness for implementation of quantum computing.
\end{abstract}


\section{Introduction}
\label{sec:Introduction}
Quantum Computing, being one of the Quantum Technologies together with Quantum Communication and Sensing, is often called a disruptive technology, i.e. has the potential to change how an industry operates in a fundamental way. Quantum Computing is supposed to resolve problems which are practically unsolvable by classical computers in cryptography, logistics and chemistry. Use of quantum computers can transform these fields, just like the introduction of classical computers changed the computation of data, in a profound way. Quantum Technology has the potential to transform human life in all kind of aspects, which makes it a technical revolution \cite{purohit2024building}.

Quantum Computing is an emerging technology which is thought of as applicable for various use cases. But at no level of its computing stack a conclusive end state is reached, i.e. from the hardware implementation of qubits to the development of algorithms and software, research still is conducted at a fundamental level. Although considerable progress has been made in the fields of hardware development, error mitigation, algorithm optimisation, software solutions, etc., many challenges still have to be overcome before quantum computers give a quantum advantage, i.e. applying quantum software which performs better than classical software giving a solution to a real-world problem. The roadmap of hardware providers is impressive but a promise \cite{Gambetta2023},\cite{Pasqal2024}. According to the annual survey done by Michiel Mosca, asking experts in the field, as time progresses chances are increasing a quantum computer being able to solve RSA-2048 within 24 hours. A majority of these experts consulted in 2023 expect this to happen within a time range of 10 to 20 years \cite{mosca2023}. 

We are now in the period called Noisy Intermediate-Scale Quantum (NISQ-)computers where the number of qubits does not exceed the hundreds and are still error-prone  \cite{preskill2018quantum}. These computers are important in their own right for experiments, learning to program quantum computers and the challenge to develop algorithms capable of making use of computers without full error-correction and an abundance of qubits. 

The uncertainty and non-linear progress of this emerging technology makes that the development and use of quantum computing within an organisation will be equally non-linear with possible parallel tracks and reversal of direction \cite{vik2021balanced}. Moreover, a step by step or level by level approach is not always expedient. This affects how organisations should act when implementing Quantum Computing and makes it more difficult to set out a roadmap for themselves to use this technology. But, being a disruptive and even revolutionary technology makes it important for organisations to be knowledgeable about the potential of quantum computing for their business, and set appropriate steps to be prepared for using quantum technology.

\section{Readiness Levels}
\label{sec:Readiness Levels}
Technology Readiness Levels (TRLs) are in use since the early seventies of the last century when NASA used this categorisation to determine whether technologies were ready to be used in space. These levels are since then slightly revised but always conveyed the message that technology, when successful, develops from fundamental research towards flawless production and use, in typical nine steps \cite{heder2017nasa}. To determine the specific level an assessment is conducted, inviting to present specific results. TRLs, which are technology centered and application specific, are often used in a general and global way, i.e. the level of readiness is being determined across laboratories, organisations and companies as a general, international indication of the state of a specific technology. TRLs as such are not the answer to the question whether an organisation or company is ready to use a specific technology to accomplish its goals. The model of nine steps to mature production has been an inspiration for various readiness levels which all describe aspects of importance for implementation of technology within a company. 


Enholm et al. \cite{enholm2022artificial} discern three categories influencing the readiness level of an organisation to deploy an AI-project: technological, organisational and environmental. In the technological category the authors refer to the availability, volume, velocity, freshness, variety and quality of data used. They deem the infrastructure, i.e. computing power, cloud infrastructure and algorithms also of importance. For a company to be ready in technical terms it is also important to interface with other systems in use. For this, one can discern System Readiness Levels and Integration Readiness Levels \cite{sauser2006trl}. Within the organisational category six aspects are considered. Whether top management is supporting the introduction of a new technology is the strongest determinant of success. When the culture within an organisation is innovative and employees are willing to learn, introduction of a new technology is much easier than within a more conservative culture. When the organisation has ample financial resources and employees have the necessary skills already or being able to learn required new skills affects the chances of success. Whether employees have confidence the new technology will perform according to the requirements, is a determinant of success. What the relative advantage of introducing a new technology will be and the compatibility, i.e. the fit between the new technology and the problem to be solved, is part of the strategy. The clearer this strategy, the more successful the introduction of the new technology will be. In the environment of the organisation ethical and moral aspects, regulations and environmental pressure, i.e. competitive and costumer pressure are to be considered.

A categorisation which describes organisational readiness not explicitly but still should be taken into account is the Market Readiness Level (MRL). A MRL estimates whether competitive products are available, what the demand for the product is, whether the customer is ready to use it and whether the product is available for widespread distribution \cite{vik2021balanced}. Purohit et al \cite{purohit2024building} describe a Quantum Commercial Readiness Level-scheme consisting of five levels, which starts with the demonstration of a QT-solution for a specific problem or need. The second level focusses on the question whether the proposed solution aligns with the needs and preferences of the potential customers. At the third level, a Minimal Viable Product is developed and with feedback from potential customers adjusted. At the fourth level alignment of the product and market is validated. The highest and fifth level is reached when scaling of production and growth of sales is seen.

Before one takes a new product to the market, it should be checked whether it complies with regulations. For emerging technologies these are not always clear because often no rules exist applicable to these products or technologies being used for production or parts of the product. Regulatory Readiness Levels determine whether regulations restrict commercialisation of the product, which is often overlooked by producers \cite{vik2021balanced}. For Legal Readiness Levels a distinction is made between the awareness phase, the prototyping phase and the implementation phase w.r.t. legal and ethical issues \cite{bruno2020technology}. While governmental organisations most often become active after companies take new technologies in production, the general public makes use of these products and deal with the consequences. These experiences are not always positive and beneficial, giving rise to discontent and critique, sometimes society wide. Societal Readiness Levels should also be determined by looking at whether the general public accepts the technology \cite{vik2021balanced}. The SRLs are in close alignment with TRLs but its focus is not only technical but much broader. At SRL 1 and 2 the researchers learn about the societal aspects of using a particular technology, at SRL 3-6 societal stakeholders such as users are involved during testing, validation and demonstrating the prototype. While SRL 7 is the final stage of the prototype, SRL 8 and 9 depict the (pre-) market launch \cite{bruno2020technology}.

Vik et al. \cite{vik2021balanced} use Organisation Readiness Levels to categorise the degree of `domestication' of the new technology, i.e. compatibility of existing technology and the new technology or how the technology is integrated in existing social and organisational practices. An organisation is at level 1 and 2 when the technology is very different from existing practice and it's not clear how to align this. At level 3 ideas are formulated about how to deal with this problem and at level 4 this is explicitly described. At level 5 a plan for transformation is formulated. An organisation is at level 6 when substantial changes are needed and at level 7 when only minor changes are still needed. At level 8 the new technology is ready to replace the existing practices which is done when an organisation is at level 9. Bruno et al \cite{bruno2020technology} present Organisation Readiness Levels as the impact of a new technology on professional roles, competences and skills, organisational functions, processes and infrastructure. At ORL 1 and 2 there is a growing awareness within the team about organisational readiness issues. In ORLs 3-6 roles, functions and structures needed to implement the technology are defined. ORL 7 is defined as the final stage of prototyping and ORL 8 and 9 are  (pre-) implementation phases where all the organisational changes are taken into account and solved. 

Implementation of a new technology within an organisation is often a cumbersome endeavour in terms of money, time and results. The development is un-even, non-linear, potentially disturbing and hard to grasp \cite{vik2021balanced}. The impact of using a new technology becomes more significant when the difference with the way things were done increases and even more so when new processes are developed. 

\section{Quantum Organisational Readiness}
\label{sec:Quantum Organisational Readiness}
Whether or not quantum computing is a technology to be used by a company, is a question only to be answered after awareness of the technology and its applicability is gathered which always will be the first step in a roadmap. What happens afterwards depends on various aspects such as the general TRL, whether the technology is already used by companies in the same domain and characteristics of the company itself. The first mover has to take a lot of hurdles, costing a lot of time and money, not to be spend by (fast) followers. The roadmap will be different for each company. Moreover, even when set out a roadmap, planning can and should be adjusted due to changed internal or external circumstances. 

\begin{table}[t]
\caption{Quantum Organisational Readiness Levels and required skills}
\begin{tabular}{|p{0.3cm}lp{2.4cm}p{3.7cm}|}
\hline
Phase&Level&Description&Required skills\\
\hline 
\parbox[t]{20mm}{\multirow{3}{*}{\rotatebox[y=-0.9cm]{90}{Implementation}}}&Level 9&Release and continuous improvement&Quantum software engineering, improvement and innovation management\\ 
\cline{2-4}
&Level 8&Full-scale implemen\-tation and scaling&Testing and quantum software engineering\\ 
\cline{2-4}
&Level 7&Pilot projects and testing&Quantum software engineering, testing, security\\ 
\hline 
\parbox[t]{2mm}{\multirow{3}{*}{\rotatebox[y=-1.1cm]{90}{Experimentation}}}&Level 6&Planning the end to end flow&Quantum software engineering, procurement, legal, human resource management\\ 
\cline{2-4}
&Level 5&Prototype development&Quantum algorithmic knowledge, quantum software engineering, human-machine interfacing\\ 
\cline{2-4}
&Level 4&Business case validation&Business development, human resource management\\ 
\hline 
\parbox[t]{2mm}{\multirow{3}{*}{\rotatebox[y=-1cm]{90}{Exploring}}}&Level 3&Strategic roadmap&Project management, education\\ 
\cline{2-4}
&Level 2&Exploratory Analysis and feasibility study&Business development, quantum algorithm engineering, quantum software engineering\\ 
\cline{2-4}
&Level 1&Awareness of Quantum Computing and its potential for the company&Basic quantum knowledge, domain knowledge, communication\\ 
\hline
\end{tabular} 
\label{tab:Quantum Readiness Levels and required skills}
\end{table}

In Table \ref{tab:Quantum Readiness Levels and required skills} we present the Quantum Organisational Readiness Levels. To create room for flexibility of what needs to be done exactly, we have arranged the readiness levels in phases. It starts with the phase of exploring quantum computing, i.e. getting to know the technology and its potential, do feasibility studies and create a roadmap to get an idea of the necessary development when opting for implementation. The second phase is experimentation, where prototypes are developed, a business case is validated and the end to end flow is planned. The third and last phase is when implementation is commenced and a production line is set-up. Each phase has three levels which are, typical but not necessarily, executed sequential.

\begin{figure}[t]
\center
\includegraphics[scale=0.33]{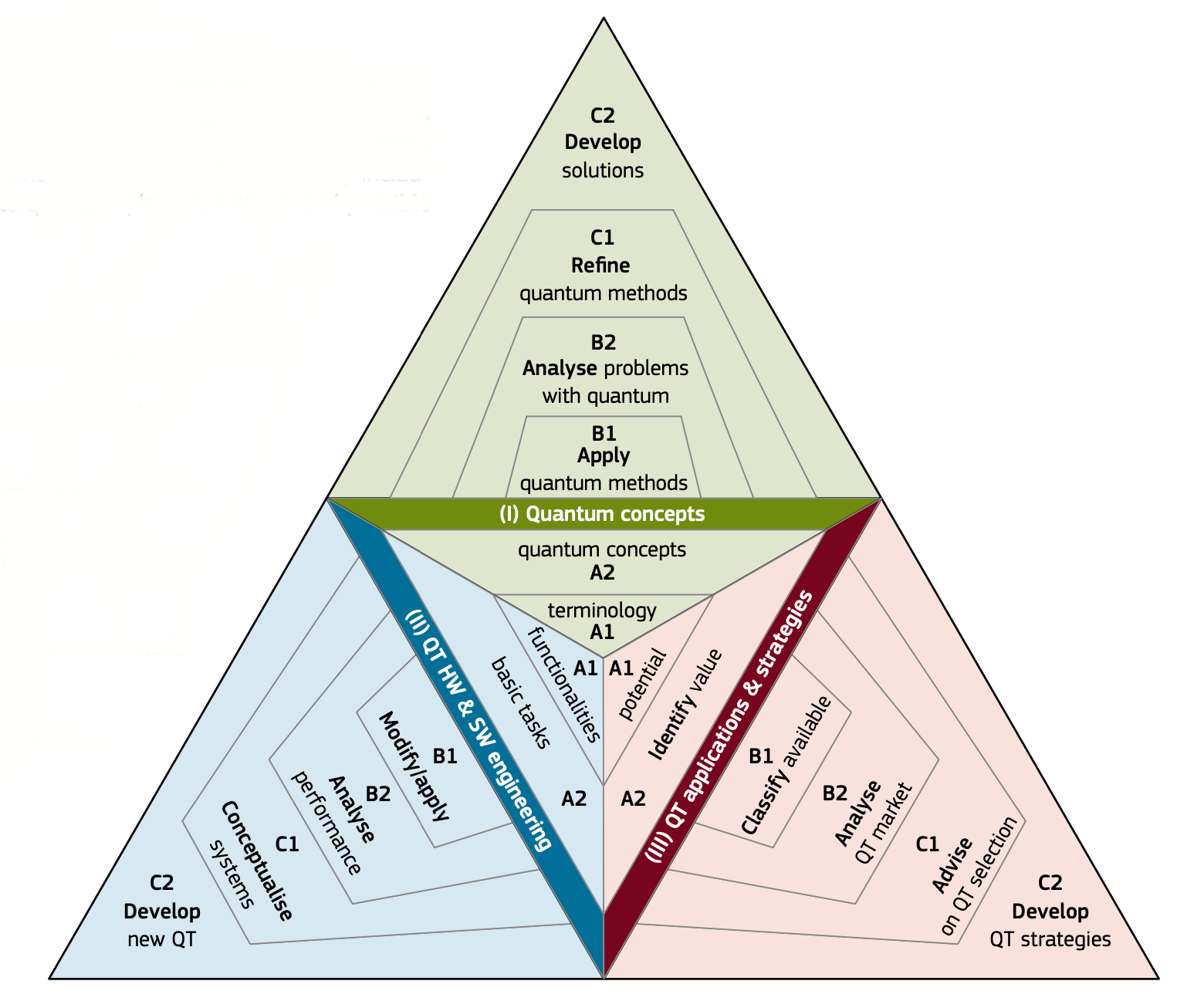} 
\caption{Competence Triangle. From: \cite{Greinert2024}}
\label{fig:Competence Triangle}
\end{figure}

Each Organisational Readiness Level is elaborated shortly, by mentioning specific activities, questions to be answered, typical jobs and corresponding competences. Job titles are traditional but some are also directly related to Quantum Technology and derived from the European Competence Framework for Quantum Technology which mentions several job profiles \cite{Greinert2024}. The competence framework discerns three fields of knowledge and skills as shown in Figure  \ref{fig:Competence Triangle}: quantum concepts in the general field of physics, quantum hardware and software engineering focussing on technology development, and applications and  strategies dealing with business opportunities. Each field can be mastered at six proficiency levels and differentiates awareness of the field (A1) up until becoming a specialist (C2), with all levels in between such as the level of bachelor (B2) and master degree (C1). Although formal education can give a head start, knowledge and skills can also be acquired and developed working with quantum technology and getting more and more experienced in the field.

\subsection{Exploring}
\label{subsec:Exploring}
During the exploring phase the company is getting aware of quantum computing and discloses the potential of the new technology. To get a feel for the technology and an initial proof of the potential, use-cases are explored and an initial roadmap is being sketched.

\subsubsection*{Level 1, Awareness of Quantum Computing and its potential for the company}
\label{subsubsec:Level 1, Awareness of Quantum Computing and its potential for the company}
Starting with no knowledge about Quantum Computing at all, the first step towards applying this emerging technology is to become aware of its potential in general and more specific within a field of business. Learning about the general principles and the revolutionary differences with classical technology is the goal during this level of readiness. The activities are: setting up a network of people and organisations in the domain of Quantum Technology, visiting conferences and attending workshops. In short: getting involved in the quantum ecosystem. 

The organisation has reached level 1 when the questions `Is it useful for my company?', `Does it make sense to implement QC in my field of business?', `Can problems, the organisation deals with, be solved more easily using quantum technology?' and `Are some companies or organisations in my field of business already investigating application of Quantum Computing?', can be answered.

To answer these questions in tentative manner, a company needs someone who is knowledgeable about Quantum Computing and also knows about the business the company is in. When Quantum Computing is indeed a technology with commercial opportunities for the company this person should act like an ambassador for Quantum Computing. To perform this task successfully, communication skills are indispensable. 

The typical job for this is most often called `Innovation Officer'. With reference to the competence framework the innovation officer should be a Quantum literate person, i.e. being in all fields at level A2. He or she becomes more and more knowledgeable when the organisation chooses to be involved in Quantum Computing.

\subsubsection*{Level 2, Exploratory Analysis and feasibility study}
\label{subsubsec:Level 2, Exploratory Analysis and feasibility study}
In this phase an exploratory analysis is conducted to identify possible use cases. When determined, these use cases should be elaborated to create a Proof of Concept, and gather some evidence on the practicality and usefulness of quantum computing. Such a Proof of Concept should preferably mimic a real situation, give a quantum solution  within an operational setting and be extendible to a prototype. When the Innovation Officer is convinced Quantum Computing promises better results than classical techniques, a preliminary feasibility study should be compiled to show if and when an implementation will be worthwhile and possible. This study should contain financial data, needed skills, organisational changes, risk management, time lines, etc. Because of the specifics of this new technology special focus should be given to the emerging and disruptive aspects of Quantum Computing. 

Related questions are: `What are the use-cases you envision?', `Did you implement a quantum algorithm and created a Proof of Concept?', `Was this application developed in-house or was it done externally?', `What are the restrictions of the application you encountered?', `Do you have a preliminary plan to implement quantum computing within your company?'

The skills needed during this phase are those of a Business Developer in close cooperation with the Innovation Officer, a Domain Expert, a Quantum Algorithm Engineer and a Quantum Software Engineer. The Business Developer should be a quantum literate overall and at level B2 in the field of applications and strategies. The Domain Expert should be at least quantum aware, i.e. in all fields at A1 level. The Quantum Algorithm Engineer and Quantum Software Engineer should be at least at level C1 in the field of quantum technology software.

\subsubsection*{Level 3, Strategic roadmap and skill development}
\label{subssubec:Level 3, Strategic roadmap and skill development}
In this phase the company is convinced Quantum Computing offers an economic potential but has to figure out how to organise the incorporation of this new technology in such a way it also gives an economic advantage. A team will be formed to develop a strategic roadmap for quantum computing integration which is aligned with existing and, possible new, organisational goals. An important part of this roadmap will be the development of the necessary quantum capabilities within the organization, i.e. forming dedicated teams for quantum projects and an infrastructure making use of the new technology. 

Questions to be answered are: `Have you defined how QC can be useful for your business?', `Is it a large or small part of your business?', `What are the time lines for development?', `How do you organise this?' and `What are the quantum skills needed?'.

The skills needed are: strategic planning, project management, quantum computing education, team building, talent acquisition and development. To create a roadmap and skill development a Product Strategist and Educational Officer are needed. The Product Strategist should be at level C2 in the field of applications and strategies and level B2 in the field of Quantum Computing. The Educational Officer should be quantum literate, i.e. at level A2 in all fields, in order to organise quantum education.

\subsection{Experimentation}
\label{subsec:Experimentation}
During this phase the company corroborates the potential of Quantum Computing by involving many more people, validating the business case, initiates the first steps to implementation and gets a grip on the end-to-end flow. This phase encompasses activities which are not irreversible as during the next phase.

\subsubsection*{Level 4, Business case validation}
\label{subsubsec:Level 4, Business case validation}
Although the top management already knows about these new developments the introduction of Quantum Computing in the company is still in its infancy. At this level the introduction of Quantum Computing evolves into a company-wide endeavour for which the approval of all concerned is needed, the business case will be validated and the techniques used will be refined. During this phase people have to be convinced the business case is worthwhile and should be pursued. An end-to-end assessment of the business case for quantum solutions, including scalability and integration challenges will be presented. To convince people, first of all they should know what it is all about and so, they should become aware of the Quantum Computing, it's differences with classical computing, the advantage of implementing, the changes needed in the company, etc. Afterwards, a decision-making process should take place. 

Questions during this phase are: `How much business value will you generate by using QC?',  'Does top management agree on the business case?', `Does the company has a long-term quantum strategy?', 'Is everybody concerned, convinced implementation is the right way?'. What is the advantage for the customer using the new product?, `Does the envisioned product aligns with regulation?'.

A job typically required for this level is the Product Strategist who should operate at level C2 in the field of strategy and applications and B2 in both the fields of concepts and technology. Because this is the level where the company chooses to continue on the path of developing a quantum technology product all concerned in this decision-making process should be at an awareness level (A2) in the field of strategy and applications.

\subsubsection*{Level 5, Prototype development}
\label{subsubsec:Level 5, Prototype development}
During this phase the company wants to build on the Proof of Concept, constructed at Level 2, and create a prototype making use of raw data and results taken from the actual use cases, including noise within the data. Algorithms tend to be developed and proven to do the job with clean data and so called toy problems, which is necessary and worthwhile in its own regard, but it doesn't proof, as such, its working in a more real although still controlled environment. Creating a business-proof prototype takes this into account and lays bare the necessary pre- and post-processing needed to get acceptable results. Also, the quantum algorithm itself possibly needs some adjustment and the results have to be analysed. The uncertainty w.r.t. the overall progress of Quantum Computing, e.g. hardware implementation, should be accounted for. So, the prototype should not yet be aligned with a specific hardware implementation and reasonable assumptions about numbers of qubits and error correction must be done.

In the end the company can answer the following questions: `Do you have a clear idea about how to take advantage of QC for a specific, well defined use case?', `Can you show how the business-proof prototype could be made scalable towards industrial production' and `Do you have a worked out example of the technique used?'. 

Skills needed for this phase are data analysis and knowledge about and skills to write quantum algorithms, Quantum hardware characterisation, Quantum Software engineering, testing and validation, and a business analysis. For this a Quantum Algorithm Engineer, a Quantum Software Engineer and a Business Analyst all at level C1 are needed.

\subsubsection*{Level 6, Planning the end to end flow}
\label{subsubsec:Level 6, Planning the end to end flow}
When the top management has accepted the business case and made resources for the implementation available, and the prototype is developed, the end-to-end work flow should be designed. During this phase a production line is being designed, with at its core the software life-cycle. The software life-cycle starts with the determination of functional and non-functional requirements and continues with design, implementation, testing and deployment of the software, which will not be different when using Quantum Computing. But in the NISQ-era, and for a long time after, Quantum Computing will be combined with classical computing as a hybrid application giving rise to design decisions w.r.t the most sensible split between classic and quantum parts. Furthermore, design patterns, i.e. standard solutions for common problems, for Quantum Computing are in development and not yet matured. Debugging and testing of quantum software is still an open research question \cite{weder2022quantum}. Using Quantum Computing as a technique for production comes with uncertainty w.r.t. hardware implementation, at this moment of time. This uncertainty should be an essential part of the life-cycle, i.e. it should be accounted for as something to re-evaluate on a regular basis \cite{Gheorghe-Pop2020}. Choice and change of a specific hardware implementation, as will the use of simulators, could have a profound impact on design and implementation of the quantum software. Pre- and post-processing of data, encoding and decoding, should be part of the work flow. An interface with a user or other machine is been designed and implemented. Whether to partner with IT-service providers, hardware and software companies and system integrators, working with big tech players or buying start-ups should be decided on. Besides these technical aspects of the end-to-end work flow, also human resources, financial and all other aspects of this implementation should be considered. 

Questions during this phase are: `Is there a clear understanding of the end-to-end flow?', 'What is the plan for implementation?' 'Is it known what kind of and how many personal the company needs?'. 

Tasks during this phase are typically done by a team of Quantum Software Engineers at various levels from B2 to C2 and Human-Machine Interface Developers at levels from B1 to C1. To find these people a Human Resource Officer is needed who should be quantum literate, i.e. at level A2 in all the fields. Furthermore, a Business Analyst at level C1 should keep a good grip on the development and project and change managers should be involved.

\subsection{Implementation}
\label{subsec:Implementation}
In this phase the use of quantum computing to create products for customers is being implemented. First as a pilot and b\`eta product, then on small scale and later full scale. When it has been released and small functional improvements are being implemented, the product is mature and evolves further.

\subsubsection*{Level 7, Pilot projects and testing}
\label{subsubsec:Level 7, Pilot projects and testing}
During this phase the designed workflow is implemented. The integration of quantum solutions and conducting pilot projects in an operational setting are important activities in this phase. In this phase they become concrete and decisions, with potential long term effects, have to be made. The organisational change, in itself, is such a long term decision. But also the choice of a specific quantum hardware platform. A decision about the hardware platform is a choice for a relationship with a hardware vendor or, more likely, a service provider who has a roadmap of its progress w.r.t. the development of the provided hardware, for a decade or longer. Choosing, after a certain period, a different hardware platform will have more impact on the company's workflow than when using classical techniques. Robust software engineering techniques, such as modelling and documentation, is in this regard even more important than usual. During this level the end-to-end workflow is tested with historic data. Results will be analysed and evaluated, adjustments will be made. Furthermore, code review and certification are organised. Necessary adjustments are documented and tests are re-aligned. 

Questions which have to be answered are: `Did you choose a specific hardware platform?', `Do you have the encoding software in place?', Do you have your quantum software life cycle in place?', `Do you need certification from a formal body and if so, do you have these certificates?', `Do you have tested all parts of the work flow?', `To what extend did the company test all possible scenarios?'. Do you have a Minimal Viable Product, interacted with potential customers and adjusted the product?

Typical jobs during this phase are Quantum Software Engineers at levels B2 to C2 and a Quantum Test Engineer at level C1. Procurement Officer, Legal Officer and Human Resource Officer. Furthermore, the Safety and Security Officer, Project manager and Change manager should be involved.

\subsubsection*{Level 8, Full-scale implementation and scaling}
\label{subsubsec:Level 8, Full-scale implementation and scaling}
When all is tested in a safe and controlled environment, a small b\`eta test is done to ensure the product is working also within a real context as envisioned. Full-scale implementation of quantum solutions are prepared and executed, organisational changes and scaling across multiple business areas are implemented. Interfaces are tested and interpretation of the resulting data by humans is been evaluated. 

Questions to be answered are: `How large was the group and people or different settings used for testing?', `What were the problems encountered?', `Did the testing show significant deviation from the expected values?'. 

Typical profiles at this phase should be Quantum Software Engineers at levels B2 to C2, a Quantum Test Engineer at level C1, Project Manager and Change Manager.

\subsubsection*{Level 9, Release and continuous improvement}
\label{subsubsec:Level 9, Release and continuous improvement}
The product is been released and available for purchase or been used as an enabling technique to produce something else. It is been used within the regular work flow without any special status. Quantum Computing is fully integrated, with a focus on continuous optimization and exploring new applications. 

Typical profiles at this phase should be Quantum Software Engineers at levels B2 to C2 and a Business Developer at level B2.

\section{Discussion}
Quantum Computing is an emerging technology with a technology stack which is not yet mature, i.e. all levels of the stack are open to fundamental change. Because the use of a different implementation at a lower level of the stack, can imply changes at a higher level, e.g. different hardware platforms give different restrictions at the programming level, a decision to work on another hardware platform could mean going back to level 5 and build a new prototype, although the company was already at level 7 and implement the end-to-end flow. This makes it difficult to create a linear roadmap and should be taken into consideration when planning the transition to quantum computing as a means of production.

The difference between level 3, creation of roadmap, and level 4, business case validation, maybe not clear. We envision the creation of a roadmap encompassing the path towards implementation of Quantum Computing as a means of production. The business case is about whether the product itself is having business value and is worthwhile pursuing. One can argue that a company first needs a business case before developing a roadmap because why should you bother making a roadmap when the business case is not profitable. Being an emerging technology this is not so clear, because the business case should also evaluate the readiness of the company for using Quantum Computing itself. This information can only be generated by having a good look at how to implement Quantum Computing, i.e. the roadmap.

\section{Conclusion}
In this article we presented research on the different levels one can discern determining the readiness of a company for the use of quantum computing. We used the well known Technology Readiness Levels and various other readiness schemes to create Quantum Organisational Readiness Levels (QORLs). The QORLs were validated by people active in the consultancy business and advising companies wanting to explore quantum computing.

\section{Future work}
We will continue the research on Quantum Organisation Readiness Levels by looking for quantifiable and measurable metrics for readiness. For example, the number of conferences been joined, knowledge of the ecosystem, workshops attended and describing competence levels of employees, to attribute a level of readiness in the Exploration phase. Within the Experimentation phase, how many and which employees being involved in setting up the business case and whether the company published any technical paper. And for the Implementation phase whether any commercial product relies on it.

\bibliographystyle{./IEEEtran}
\bibliography{IEEEabrv,./business}

\end{document}